\documentclass[12pt,preprint]{emulateapj}   

\def\sun{\hbox{$\odot$}}

\begin{document}   

\title{MOND and the Universal Rotation Curve: similar phenomenologies}   

\author{Gianfranco Gentile
       \altaffilmark{1,2}
       }

\altaffiltext{1}{Sterrenkundig Observatorium, Ghent University, Krijgslaan 281, S9,
B-9000 Ghent, Belgium}
\altaffiltext{2}{University of New Mexico, Department of Physics and Astronomy, 800 Yale   
Blvd NE, Albuquerque, NM 87131, USA}

\begin{abstract}
The Modified Newtonian Dynamics (MOND) and the Universal Rotation Curve (URC) 
are two ways to describe the general 
properties of rotation curves, with very different approaches concerning
dark matter and gravity.  
Phenomenological similarities between the two approaches are studied 
by looking for properties predicted in one
framework that are also reproducible in the other one.

First, we looked for the analogous of the  
URC within the MOND framework. 
Modifying in an observationally-based way the baryonic contribution $V_{\rm
bar}$ to the rotation curve predicted by the URC, and applying the MOND 
formulas to this $V_{\rm bar}$, leads to a ``MOND URC'' whose properties are
remarkably similar to the URC.

Second, it is shown that the URC predicts a tight mass discrepancy - acceleration
relation, which is a natural outcome of MOND. With the choice of $V_{\rm bar}$ that 
minimises the differences
between the URC and the ``MOND URC'' the relation is almost identical to 
the observational one.

This similarity between the observational properties of MOND and the URC has no 
implications about the validity of MOND as a theory of gravity,
but it shows that it can reproduce in detail the phenomenology of disk galaxies'
rotation curves, as described by the URC.
MOND and the URC, even though they are based on totally different assumptions,
are found to have very similar behaviours and to be able to reproduce
each other's properties fairly well, even with the simple assumptions
made on the luminosity dependence of the baryonic contribution to 
the rotation curve.
\end{abstract}





\section{Introduction}

Rotation curves of spiral galaxies have been studied for several decades
now, and they are a useful tool to show the mass discrepancy
in galaxies: the observed kinematics and that predicted from the 
observed baryonic distribution do not match.
Either an additional mass component or a modification of gravity are needed.
In the recent past, rotation curves have been exploited to test
the predictions of cosmological models of structure formation in the
Universe, such as the currently favoured $\Lambda$ Cold Dark Matter ($\Lambda$CDM). 
Most observations show that the baryons are the main kinematic
component in the inner parts and that the so-called ``cuspy'' halos predicted
by $\Lambda$CDM (Navarro, Frenk \& White 1996, Moore et al. 1999, Navarro et
al. 2004) fail to reproduce observed rotation curves
(see e.g. de Blok et al. 2001; de Blok \& Bosma 2002; Gentile et al. 2004, 
2005, 2007a,b; McGaugh et al. 2007).

Rotation curves have been shown to have some general properties that
can be described by a stellar disk + dark matter halo model (the
Universal Rotation Curve, URC; Persic, Salucci \& Stel 1996, hereafter PSS). 
In this framework, the circular velocity of a spiral galaxy at a 
certain radius depends only on one parameter, e.g. the total luminosity
of the stellar disk.

MOND, the Modified Newtonian Dynamics (Milgrom 1983) is another successful 
prescription to predict the rotation curve
of a spiral galaxy. According to MOND, the mass discrepancy is not due
to an unseen mass component (the dark matter halo), but it is instead
the signature of the failure of Newtonian gravity to describe the 
observed kinematics at the low accelerations found in the outer parts
of galaxies. In MOND, below a certain
critical acceleration $a_0$ the Newtonian gravity is no longer valid.
Earlier concerns about the inconsistency of MOND with General Relativity
are now overcome with the TeVeS theory (Bekenstein 2004).
MOND has a remarkable predictive power for the kinematics of 
galaxies: it fits the kinematics of small dwarf 
galaxies (Gentile et al. 2007a,c), of the Milky Way (Famaey \&
Binney 2005), of early-type spiral galaxies (Sanders \& Noordermeer
2007), of massive ellipticals (Milgrom \& Sanders 2003), and it
naturally explains observed tight scaling relations in spiral
galaxies (McGaugh 2004, 2005).

\section{MOND}

The Modified Newtonian Dynamics (Milgrom 1983, see Sanders \& McGaugh 2002
for a review) can be invoked as an alternative to dark matter to explain
the observed kinematics of disk galaxies. Within the MOND framework, the 
true gravitational acceleration $g$ is linked to the Newtonian one $g_{\rm N}$ 
through the following relation:

\begin{equation}   
g=\frac{g_{\rm N}}{\mu(g/a_0)}   
\end{equation}   
\label{mondeq}
   
where $\mu(x)$ is an interpolation function whose asymptotic values are $\mu(x)=1$ when 
$g \gg a_0$ and $\mu(x)=g/a_0$ when $g \ll a_0$. $a_0$ is the critical acceleration
below which the Newtonian gravity is no longer valid; previous studies 
(Begeman, Broeils \& Sanders 1991) found that $a_0 \sim 1.21 \times 10^{-8}$
cm s$^{-2}$.
Even though in a general case a modified version of the Poisson equation
should be solved, eq. \ref{mondeq} can be shown to be a good approximation
for axisymmetric disks (Brada \& Milgrom 1995).
The interpolation function has been given usually the following functional form:

\begin{equation}   
\mu_{\rm orig}(x)=\frac{x}{\sqrt{1+x^2}}   
\label{mond}   
\end{equation}   
   
However, it is obvious that a whole family of functions are compatible with
the required asymptotic behaviours. For instance, Famaey \& Binney (2005) 
proposed that the form:

\begin{equation}   
\mu_{\rm FB}(x)=\frac{x}{1+x}    
\label{mondnew}   
\end{equation}   

could be a better choice, since, contrary to eq. \ref{mond}, it is compatible
with the relativistic theory of MOND (TeVeS) put forward by Bekenstein (2004).
Famaey et al. (2007) showed that using eq. \ref{mondnew} leads to a slightly
different value of $a_0$: $a_0 = 1.35 \times 10^{-8}$ cm s$^{-2}$.

If eq. \ref{mond} is used as the interpolation function, then 
within the MOND framework the circular velocity
velocity $V_{\rm obs}(r)$ can be expressed as a function of 
$a_0$ and the Newtonian baryonic
contribution to the rotation curve $V_{\rm bar}(r)$ at radius $r$:

\begin{equation}
V_{\rm obs}^2(r)=V_{\rm bar}^2(r)+V_{\rm bar}^2(r) 
\left(\sqrt{\frac{1+\sqrt{1+\left(\frac{2ra_0}{V_{\rm bar}^2(r)}\right)^2}}{2}}-1\right)
\label{vmond}
\end{equation}

where $V_{\rm bar}(r)=\sqrt{V_{\rm stars}^2(r)+V_{\rm gas}^2(r)}$
(ignoring the contribution of the bulge),
$V_{\rm stars}(r)$ and $V_{\rm gas}(r)$ are
the Newtonian
contributions to the rotation curve of the stellar and gaseous disks,
respectively (see Milgrom 1983).
The amplitude of $V_{\rm stars}(r)$ (whose shape is fixed by
photometric observations), can be scaled according to the chosen, or fitted, 
stellar mass-to-light ($M/L$) ratio. $V_{\rm gas}(r)$ is derived from
HI observations, when they are available.
Note that the second term of the right-hand side of eq. \ref {vmond} acts as a 
``pseudo-dark matter halo'' term and that it is completely determined
by the baryonic terms. 

If eq. \ref{mondnew} is chosen as the interpolation function
instead of eq. \ref{mond}, the equivalent 
of eq. \ref{vmond} becomes:

\begin{equation}
V_{\rm obs}^2(r)=V_{\rm bar}^2(r)+V_{\rm bar}^2(r) 
\left(\frac{\sqrt{1+\frac{4 a_0 r}{V_{\rm bar}^2(r)}}-1}{2}\right)
\label{vmondnew}
\end{equation}

(see e.g. Richtler et al. 2008). 
As expected, in both eqs. \ref{vmond} and \ref{vmondnew} the 
``pseudo-dark matter halo'' term vanishes in the limit $a_0 \rightarrow 0$.
We note MOND has a remarkable 
predictive power for the general properties of rotation curves,
and in many cases it is able to fit observed individual rotation curves
(Kent 1987; Milgrom 1988;
Begeman, Broeils \& Sanders 1991; Sanders 1996; de Blok \& McGaugh
1998) . MOND correctly predicts general scaling relations linked
to rotation curves, such as the baryonic Tully-Fisher
relation (see McGaugh 2005, even though we note that other authors
find different slopes, mainly due to different choices of the stellar
$M/L$ ratio) or the mass discrepancy-acceleration relation
(McGaugh 2004).

\section{The Universal Rotation Curve (URC)}
\label{section_urc}

\begin{figure}   
\includegraphics[width=8.5cm]{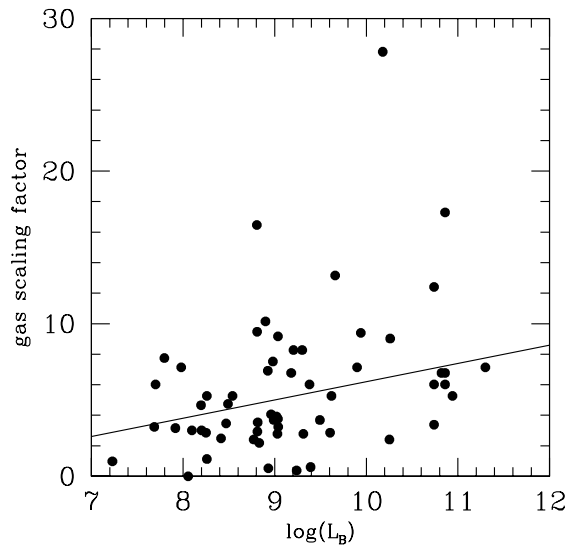}  
\caption{
Gas scaling factor versus (log of) B-band luminosity.
The data are taken from Swaters (1999) and Hoekstra et al. (2001),
and the solid line is the result of a linear fit to the points.
}
\label{scaling}   
\end{figure}

\begin{figure*}   
\includegraphics[width=17cm]{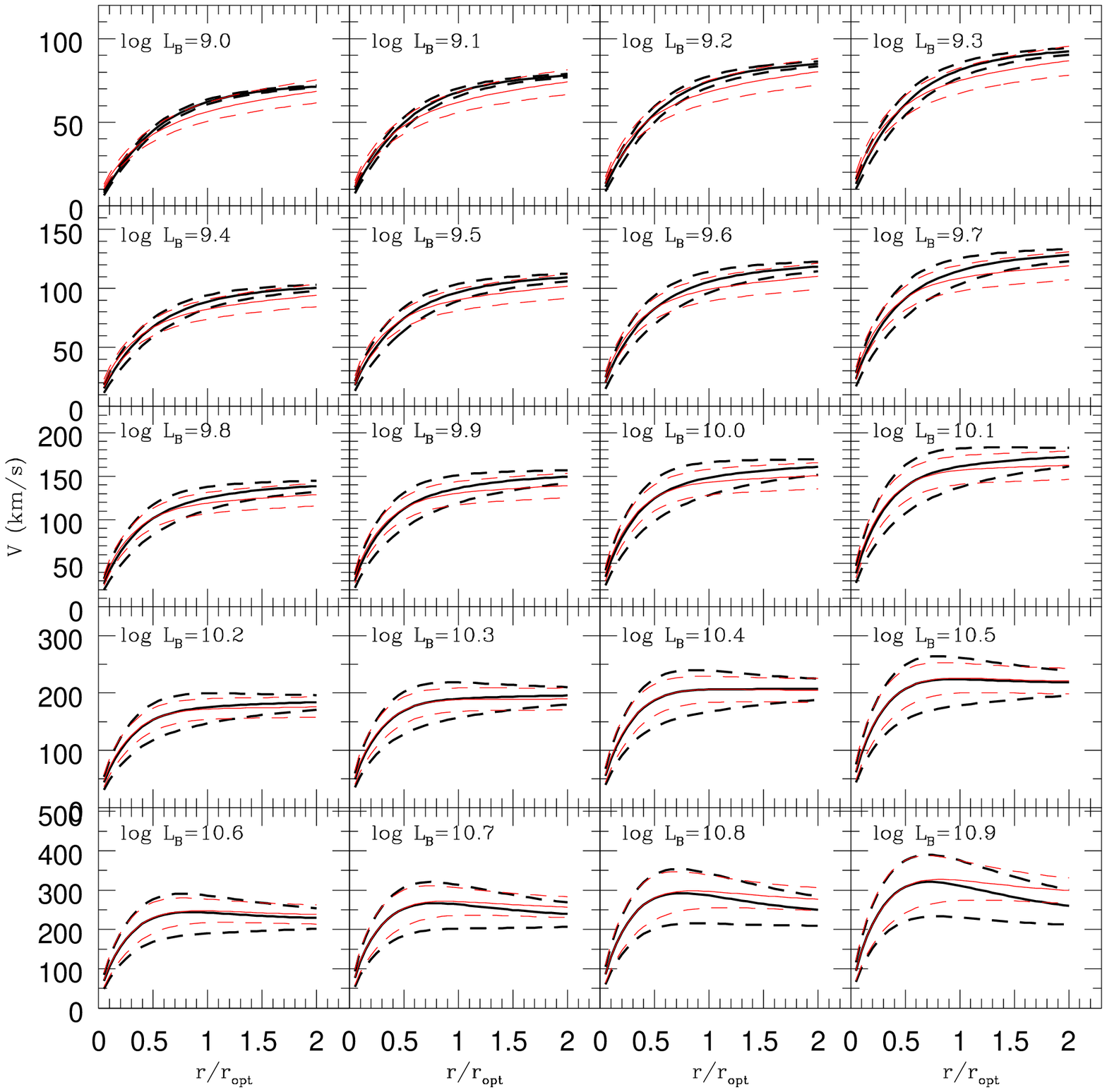}   
\caption{
Comparison between the Universal Rotation Curve (thick black curves),
at various luminosities, and the 
equivalent for MOND (thin red curves, using $\mu_{\rm orig}(x)$), 
with the stars and gas contributions defined in the text.
Dotted curves indicate the uncertainties.
}
\label{urc}   
\end{figure*}   

\begin{figure*}   
\includegraphics[width=17cm]{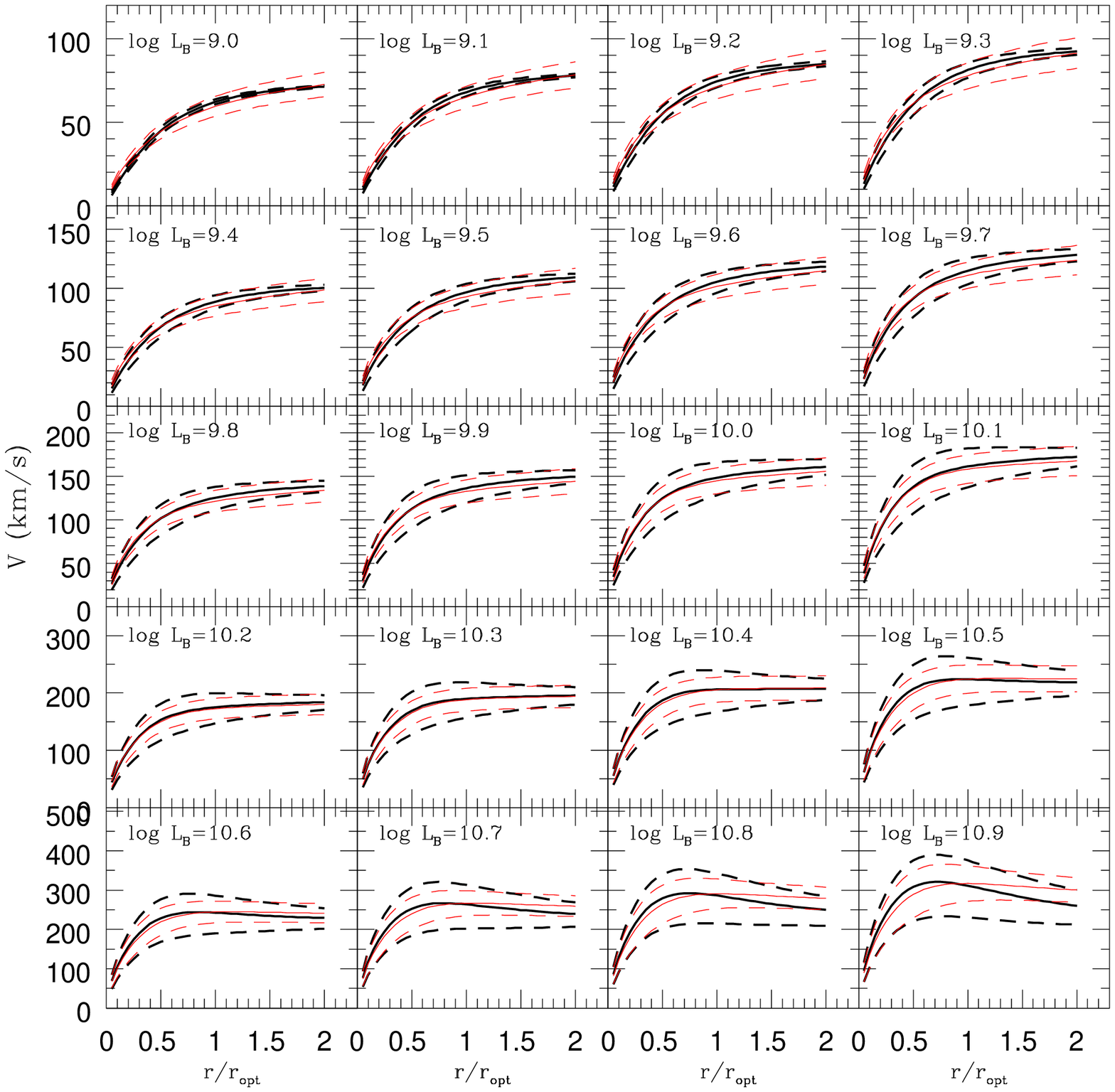}   
\caption{
Comparison between the Universal Rotation Curve (thick black curves),
at various luminosities, and the 
equivalent for MOND (thin red curves, using $\mu_{\rm FB}(x)$), 
with the stars and gas contributions defined in the text.
Dotted curves indicate the uncertainties.
}
\label{urc_new}   
\end{figure*}

PSS, from the analysis of a sample 
of more than 1000 rotation curves, showed that the circular velocity of 
a disk galaxies at a given radius (note that they considered rotation curves extended
up to 2 optical radii $r_{\rm opt}$) can be described as a function of
only one parameter, e.g. the total luminosity. The function which describes the rotation
velocity depending only on radius and luminosity is given the name Universal
Rotation Curve (URC). We note that the kinematics of real galaxies 
depend on other factors too (such as the surface brightness), but
the general phenomenological properties of rotation curves can be
described well with the luminosity as only parameter.

In detail, the circular velocity $V_{\rm URC}(r)$ at radius $r$ can be expressed
as a function of $V_{\rm opt}$ (the circular velocity at $r_{\rm opt}$), and of the  
stellar and dark contributions $V_{\rm disk, N}$ and $V_{\rm halo, N}$
normalised at $V_{\rm opt}$ (i.e., $V_{\rm disk, N}(r_{\rm opt})+V_{\rm halo, N}(r_{\rm opt})
=1$):

\begin{equation}
V_{\rm URC}(r)=V_{\rm opt}\sqrt{V_{\rm disk, N}^2(r)+V_{\rm halo, N}^2(r)}
\label{vurc}
\end{equation}

$V_{\rm opt}$, $V_{\rm disk, N}$, and $V_{\rm halo, N}$ can be in turn 
expressed as functions of $r$, $\lambda=L/L_*$ ($L$ is the total blue
luminosity of the galaxy in question and log $L_*$=10.4 in solar units), and 
$r_{\rm opt} \simeq 13 \lambda ^ {0.5}$ kpc:

\begin{equation}
V_{\rm opt}=\frac{200~\lambda^{0.41}}{\left[0.80+0.49~{\rm log}\lambda
+\frac{0.75~{\rm exp}(-0.4 \lambda)}{0.47+2.25 \lambda^{0.4}}\right]^{1/2}} ~{\rm km~s}^{-1}
\label{vopt}
\end{equation}

\begin{equation}
V^2_{\rm disk, N}=(0.72+0.44~{\rm log} \lambda) \frac{1.97(r/r_{\rm opt})^{1.22}}
{[(r/r_{\rm opt})^2+0.61]^{1.43}}
\label{vdiskn}
\end{equation}

\begin{equation}
V^2_{\rm halo, N}=1.6~{\rm exp}(-0.4 \lambda)\frac{(r/r_{\rm opt})^2}{(r/r_{\rm
opt})^2+2.25 \lambda^{0.4}}
\label{vhalon}
\end{equation}

According to PSS, eqs. \ref{vurc} to \ref{vhalon} predict the circular velocity of a spiral 
galaxy at radius $r$ with an uncertainty of 4\%. 
Choosing $r_{\rm opt} \simeq 13 \lambda ^ {0.5}$ kpc 
hides any variation of surface brightness for a given luminosity, which
results in different rotation curve shapes for galaxies with different
surface brightnesses (see e.g. McGaugh \& de Blok 1998). 
However, we note that the original formulation of 
the URC (with the radius $r$ in units of the stellar exponential scale length $r_{\rm D}$) 
takes this effect into account, since the rotation curves of low- and high-surface 
brightness galaxies are virtually indistinguishable once the radius 
is expressed in units of $r_{\rm D}$ (e.g., Verheijen \& de Blok 1999). 
Also, in the URC the contribution of the gaseous disk is ignored. This is 
acceptable in the range of luminosities considered by PSS because a) the
gasous disk never dominates the kinematics and b) because of its scaling 
with the dark matter contribution, its contribution can be implicitely accounted
by $V_{\rm halo}$.

An important feature of the URC is the strong dependence of the logarithmic
slope of the rotation curve $\nabla$ on the luminosity.
They define $\nabla$ in the range $0.5r_{\rm opt} < r < r_{\rm opt}$: 
$\nabla=d{\rm log}V/d{\rm log}r|_{0.5r_{\rm opt} < r < r_{\rm opt}}$.
Fainter galaxies have an increasing rotation
curve around $r_{\rm opt}$, while in the most massive galaxies $V(r)$ is already
decreasing around $r_{\rm opt}$.

\section{The MOND Universal Rotation Curve}
\label{MURC}

The question that this paper is addressing is: since both MOND and the URC
are successful ways to predict the general kinematical properties of disk 
galaxies, to what extent do they have similar properties?
One way to compare their properties is to try to derive something like
the URC within the framework of MOND. In order to achieve this, both
$V_{\rm stars}(r)$ and $V_{\rm gas}(r)$ are needed. 
For the former term, the URC provides us with
a recipe to estimate it, while the latter is ignored in the URC.
A rough estimate of $V_{\rm gas}(r)$ can be made through the scaling property
of $V_{\rm halo}(r)$ compared to $V_{\rm gas}(r)$ noticed by e.g. Bosma (1981) and
Hoekstra, van Albada \& Sancisi (2001): $V^2_{\rm halo}(r) \approx n 
\times V^2_{\rm gas}(r)$.
The factor $n$ 
was calculated by performing a linear fit to the observational data 
in the scaling factor versus luminosity plot (see Fig. \ref{scaling}). 
The data were taken from the 
samples of Swaters (1999) and Hoekstra et al. (2001). The former work 
gives R-band luminosities, whereas the URC uses B-band. The B-band luminosities
of the Swaters (1999) sample were calculated from the LEDA database, when available,
and from NED. The linear fit gives $n = 1.3 {\rm log} \lambda - 6.4$.
Therefore, we estimated $V_{\rm gas}(r)$ through the following
relation: $V_{\rm gas}(r) \sim V_{\rm halo,URC}(r)/\sqrt{1.3 {\rm log} \lambda - 6.4}$.
This turns out to be in line with the findings of Swaters (1999),
who finds $n$ to be correlated with surface brightness, and hence with luminosity.
Such a scaling of the actual surface density with the gas surface density
would happen in the case of a conspicuous amount of disk dark matter,
e.g. H$_2$ in the form of clumps (Pfenniger, Combes \& Martinet 1994)
or in a cold neutral medium phase (Papadopoulos, Thi \& Viti 2002).

The MOND formulas (eqs. \ref{vmond} and \ref{vmondnew}) were then applied to
the $V_{\rm stars}(r)$ and $V_{\rm gas}(r)$ derived above.
However, using the face value $V_{\rm stars}(r)$ coming from the URC 
formulas leads to a MOND Universal Rotation Curve that is quite 
different from the original URC, especially for galaxies where 
the stellar disk dominates the kinematics. In the lower panels
of Figs. \ref{urc} and \ref{urc_new} 
the MOND curves would lie significantly above the URC curves.
Thus, a parameter $\eta$ such that $V_{\rm stars, MOND}^2(r)=
\eta V_{\rm stars, URC}^2(r)$ was fitted in order to have the best possible
agreement between the URC and the "MOND URC", 
assuming that the scaling of the stellar mass-to-light ratio with luminosity
predicted by the URC is correct. 

This parameter $\eta$ corresponds to the ratio of the stellar M/L ratios
required in MOND and the URC frameworks, respectively.
A best-fit value of $\eta=0.77$ was found for the standard interpolation
function (eq. \ref{mond}) and a value of $\eta=0.57$ for the simple one 
(eq. \ref{mondnew}). This is in agreement with Famaey et al. (2007), who
compare rotation curve fits made with the two $\mu$ functions considered
here are conclude that the simple $\mu$ function gives stellar $M/L$ ratios
that are on average $\sim$30\% lower than with the standard $\mu$.

The stellar mass-to-light ratios are generally not well known in spiral 
galaxies. Here we considered a conservative estimate of the uncertainties
to be 50\% (see de Jong \& Bell 2006). The uncertainty on the gas scaling
factor $n$ was taken to be 3.4, from the scatter in the distribution of
$n$ shown in Fig. \ref{scaling}. 
This scatter is large, but it is not possible to find more a accurate
way to estimate the scaling factor within the context of the URC, because 
one would need to consider other parameters (e.g. the Hubble type) than 
the luminosity, which are much more difficult to take into account: for 
instance, in early-type disk galaxies usually there is a central depression 
in the HI distribution, but its size and magnitude are variable. Moreover,
the URC, by construction, has the luminosity as the only parameter
varying from galaxy to galaxy. 
The uncertainties on the baryonic contribution
to the rotation curve result in uncertainties on the calculated rotation curves.
They are only rough estimates, since other (less easily quantifiable) sources of 
errors were ignored, such as the presence of the bulge.
These uncertainties on the stellar $M/L$ ratios mean that the disagreement
between the MOND and URC stellar $M/L$ ratios (i.e., $\eta$ being $
\neq 1$) does not necessarily imply that the approach of the present paper
is incorrect. Moreover, this is in qualitative agreement with the fact that 
1) in McGaugh (2005) the ratio of the ``maximum'' $M/L$ ratio and the 
MOND $M/L$ ratio has a median value of 1.8, where ``maximum'' in McGaugh (2005)
is admittedly loosely defined but it is generally taken as a fit where the peak 
of the rotation curve can be entirely explained by the stellar disk; and 2)
the URC has disks close to ``maximum'' (following the above definition) for most
galaxies apart from the least massive ones.

The URC and the ``MOND URC'' (using $\mu_{\rm orig}(x)$)
are compared in Fig. \ref{urc}: for the whole
range of circular velocities and radii usually sampled by rotation curves,
the two datasets display almost identical properties. 
Not surprisingly, the best agreement is found
for the range $100~{\rm km~s^{-1}} \lesssim V_{\rm opt} \lesssim 250~{\rm km~
s^{-1}}$, where most observed rotation curves lie. 
The worst disagreement is found for the largest distances, in particular in the most massive 
galaxies. This has to be expected, though, as the number of data points in this
region of the parameter space is quite small: the sample of PSS has very few 
data for large radii of massive galaxies. Indeed, only a minority of the 
curves come from HI observations (which trace the kinematics typically out to 
2-3 $r_{\rm opt}$), and few galaxies are early-type large spirals (only
two galaxies in their sample have type Sab or earlier).  
However, using the uncertainties on the baryonic contribution discussed 
above, the URC and the MOND URC are always in agreement with each other
within the errorbars.
Also, note that in this paper we ignored the contribution of the bulge,
since it was ignored in the URC. This is likely to have a strong effect on
the rotation curves of the largest galaxies.

The same considerations can be made if one uses $\mu_{\rm BF}(x)$ instead of
$\mu_{\rm orig}(x)$. Fig. \ref{urc_new} shows that the URC and the 
``MOND URC'' using $\mu_{\rm BF}(x)$ agree
with each other as well as if one uses $\mu_{\rm orig}(x)$. 

\begin{figure}   
\includegraphics[width=8.5cm]{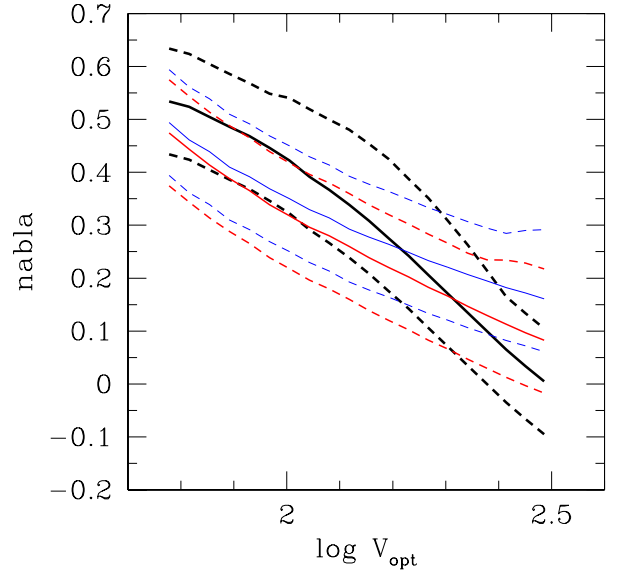}  
\caption{
Logarithmic slope of the rotation curve versus log($V_{\rm opt}$).
The thick black lines represents the URC, the (less thick) red and (thin) blue lines represent
the MOND curves, with the standard and simple $\mu$ functions, respectively.  
}
\label{nabla}   
\end{figure}

Using either interpolation function, the trend in logarithmic
slope (see Section \ref{section_urc}) is also reproduced: in the outer
parts of the galaxy, the faintest galaxies have a rising rotation curve 
while the brightest galaxies have a declining rotation curve.
This is illustrated in Fig. \ref{nabla}, where the logarithmic slope
as a function of $V_{\rm opt}$ is displayed. Again, within the uncertainties
the two formalisms agree with each other. MOND does not reproduce particularly
well the slopes of the most massive galaxies, mainly because of the fact
that the bulge is not taken into account. For consistency with PSS 
we first performed a linear fit of the rotation curves between
0.5 and 1 $r_{\rm opt}$, then we took the slope at 1 $r_{\rm opt}$.
The uncertainties on the slopes were computed as the maximum between 
the propagation of the uncertainties on the rotation curves and 
a minimum value of 0.1 (see PSS).

\section{The mass discrepancy-acceleration relation for the URC}

Now that it has been shown that MOND can reproduce something which is
quite close to the URC, in order to further compare the two phenomena
let us have a look if the URC can reproduce one of the main properties linked
to MOND: the mass discrepancy-acceleration relation.
McGaugh (2004) showed that rotation curve data seem to organise themselves
along a relation linking the mass discrepancy (defined as $D(r)=V^2(r)/V^2_{\rm
bar}(r)$, where $V(r)$ is the rotation velocity at radius $r$) to the
gravitational acceleration (here the Newtonian acceleration is used,
$g_{\rm N}=V^2_{\rm bar}/r$). In the same paper, it is also shown that
the stellar $M/L$ ratios arising from the MOND fits minimise the scatter
in the mass discrepancy-acceleration relation.

\begin{figure*}   
\includegraphics[width=17.5cm]{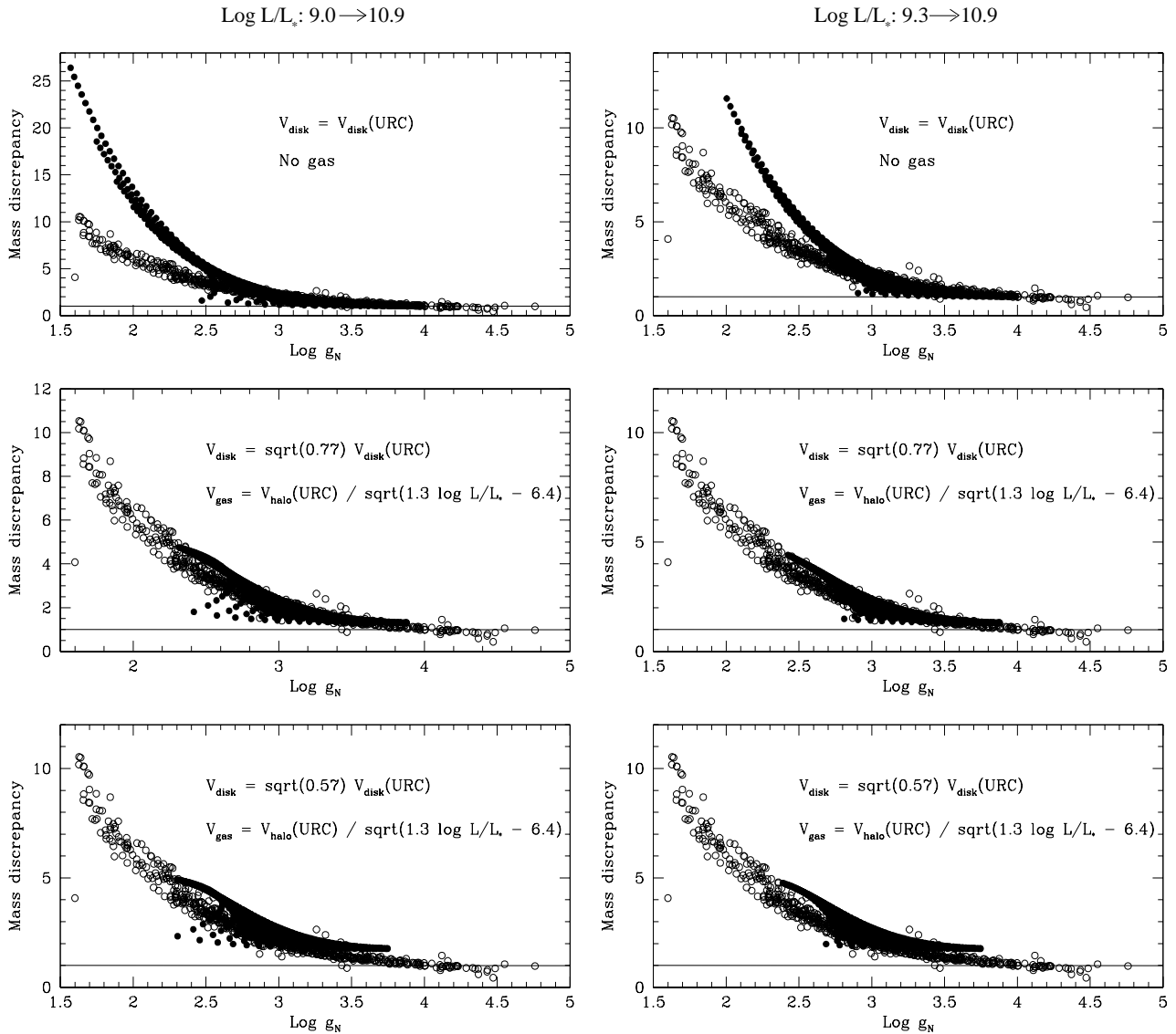}   
\caption{
Full circles: mass discrepancy-acceleration (MDA) 
relation using the URC formulas.
Open circles: the MDA data from McGaugh (2004), using the stellar
M/L ratios arising from the MOND rotation curve fits.
The horizontal axis is the log of the Newtonian acceleration (in units of km$^2$ s$^{-2}$
kpc$^{-1}$) and the mass discrepancy is defined with respect to the 
URC rotation curve.
The ranges of B-band luminosities are indicated at the top (in solar units)
and the radii range from 0.05$r_{\rm opt}$ to 2$r_{\rm opt}$, in steps
of 0.05$r_{\rm opt}$.
In the upper panel the face-value URC prescription for the baryons
was considered, while for the middle and bottom panel the 
values for $V_{\rm disk}$ and $V_{\rm gas}$ were taken as in
Section \ref{MURC}.
}
\label{mda}   
\end{figure*}

Fig. \ref{mda} shows that using the URC formulas one finds 
such a relation, whatever choice of baryonic mass is made.
Since we are looking for such a relation in the URC context, the mass 
discrepancy is always computed with respect to the URC rotation curve
(i.e., in the above definition of $D(r)$, $V^2(r) = V^2_{\rm URC}(r)$).
Obviously, using $V^2(r) = V^2_{\rm MOND}(r)$ (whichever $\mu$ is used)
leads to a zero-scatter relation, by construction.
A priori the tightness of the relations shown in Fig. \ref{mda}
comes a bit as a surprise: while in MOND a critical acceleration
is inbuilt in the theory, in the URC framework there is no such thing,
at least in an explicit way. Semi-analytical dark matter models 
(e.g., van den Bosch \& Dalcanton 2000) can also reproduce
a mass discrepancy-acceleration relation using the right choice of parameters.
However, the main thing to note is that in the present paper the URC gives a 
mass discrepancy-acceleration
relation without any parameter to adjust.

From the lower four panels Fig. \ref{mda} one can see that the mass 
discrepancy-acceleration relation from the URC formulas
is consistent with that found by McGaugh (2004) with observational data
if the baryonic contribution to the rotation curve is chosen to be the same
as in Section \ref{MURC}, with $\eta=0.57$ or 0.77. 
The former value of $\eta$ slightly overestimates the mass discrepancy
(compared to McGaugh's points) for a given Newtonian acceleration
$g_{\rm N}$: this is to be expected because $\eta=0.57$
corresponds to the ``simple'' interpolation function, where the full MOND
regime is reached faster than with the ``standard'' function.
Using $\eta=0.77$ gives
an excellent agreement: it corresponds to the ``standard''
interpolation function, which is what McGaugh used. 
Taking the face-value $V_{\rm bar}(r)$ predicted by the URC yields a remarkably
tight relation; the scaling is obviously different, since the stellar $M/L$
ratio is different and the gaseous contribution is ignored.
We also considered different ranges of luminosities:
since the URC is defined for $V_{\rm opt} \gtrsim 80$ km s$^{-1}$
(see PSS), the predicted scaling relations are not expected to hold
with high accuracy. Indeed, the left panels Fig. \ref{mda} show that 
for $V_{\rm opt} \lesssim 80$ km s$^{-1}$ the innermost few points of the 
rotation curves start to become discrepant.

\section{Discussion}

One of the main differences between the URC and the MOND fits
is the stellar $M/L$ ratio: while 
in the former disks are mostly close to maximum (following the definition
in Section \ref{MURC}), the latter yields slightly submaximal
disks (see McGaugh 2005), which results in a very different scaling in the mass discrepancy - 
acceleration plane. 
Again, we note that a limitation of the present approach is that these statements 
ignore any surface brightness variation at equal luminosity.
A robust, precise and incontrovertible method to determine 
stellar $M/L$ ratios is yet to be found. A number of methods have been suggested
(see e.g. de Jong \& Bell 2006), which suggest $M/L$ ratios ranging roughly between 
half-maximum to maximum disks. Therefore, a distinction between the URC 
and MOND based on the predicted stellar $M/L$ ratios is not viable at the moment.
The fact that the simple $\mu$ function gives lower stellar $M/L$ 
than the standard one has
already been shown by Famaey et al. (2007).

In the present paper the discussion is focussed only on the 
general properties of spiral galaxies' rotation curves. Both MOND and the
URC are successful ways of describing the overall behaviour of rotation 
curves, but inevitably there are cases where they fail to reproduce the
details in the rotation curves. Also, only a rough estimate of $V_{\rm bar}(r)$
was made in the context of MOND, in particular for $V_{\rm gas}(r)$. 
The observed $V_{\rm halo}(r)$/$V_{\rm gas}(r)$ scaling shows considerable variations
around the value of $\sqrt{1.3 {\rm log} \lambda - 6.4}$ that was assumed here. 
Despite these large approximations,
the URC and MOND exhibit remarkable phenomenological similarities.

A major feature of MOND and the URC is that in both frameworks one is able
to infer in a reasonably accurate way the general properties of rotation 
curves based on the amount and
distribution of baryonic matter only. This is closely linked to another
property of rotation curves (Sancisi 2003), whereby for any feature in a
rotation curve there is a corresponding feature in the baryonic distribution.
The literature is full of such examples, one of the most spectacular being
NGC 1560 (Broeils 1992). This empirical fact is unexpected in a Cold Dark
Matter (CDM) framework since CDM fits to the rotation curves require strongly
submaximal disks, which would cancel this baryon-kinematics coupling that
is observed (see e.g. de Blok et al. 2001, McGaugh 2004, Gentile et al. 2004).
Yet another evidence of the close coupling of baryonic
and general properties of spiral galaxies is the observed correlation between stellar
exponential scale lengths and halo core radii (Donato, Gentile \& Salucci
2004). A similar effect can be inferred also within MOND (Milgrom \& Sanders 2005).
The interpretation of the baryon-kinematics coupling is not obvious: possible
solutions are that baryons and dark matter are interacting in some unknown way
other than gravity in the
central parts of galaxies or that a MOND-like 
theory such as TeVeS is the correct theory of gravity.

A possible concern about the approach presented in this paper might have been that
the similarity between MOND and the URC might actually be an identity.
This is not the case, since even if one assumes the same baryonic
contribution in both frameworks (e.g. the ones considered in Section \ref{MURC}, with either
$\eta=0.57$ or 0.77), the functional form of the halo contribution in the
URC context (eq. \ref{vhalon}) is completely different from the functional
form of the ``pseudo-halo'' contribution in MOND (second term of the
right-hand side of eq. \ref{vmond}).
 

As pointed out by Salucci \& Gentile (2006), any alternative theory of gravity
(such as MOND or MOG, see Moffat 2006)
should be able to account for the observed phenomenology of rotation curves. 
In the present paper we have demonstrated that MOND does fairly well, despite
the naive expectation of having only flat rotation curves because of the 
asymptotically flat behaviour of the MOND rotation curves. Hence,
the MOND framework not only usually gives good fits to rotation
curves, but it also predicts the correct scaling of the rotation curves
properties with luminosity, as shown by the URC.

\section{Conclusions}

Starting from the consideration that both MOND and the Universal Rotation
Curve (URC) are valid ways to describe the general properties of rotation curves,
the two approaches were compared by trying to reproduce one's predictions
using the prescriptions of the other one.

The first comparison was made by attempting to reproduce something like
the URC in the MOND framework. Some (observationally-based) assumptions
on the stellar and gaseous contributions to the rotation curve were made.
It turns out that it is possible to build a ``MOND URC'' with very similar
properties to the URC. 
The MOND URC also exhibits the trend in logarithmic slope
versus luminosity seen in the URC.
Both the ``standard'' MOND interpolation function and the simple one proposed 
by Famaey \& Binney (2005) were tested, with almost identical results.

The second comparison was made by looking for a mass discrepancy -
acceleration relation using the URC formulas. 
With all three choices 
of baryonic contributions ($V_{\rm bar}$) considered in this paper a tight (unexpected
a priori) relation arises. Using the $V_{\rm bar}$ values that match best the
``MOND URC'' to the URC one finds a mass discrepancy - acceleration relation 
like the observed one.

While these results have no implications as to whether MOND is a valid
theory of gravity,
MOND and the URC, even though they are based on totally different assumptions,
are found to display similar properties and to reproduce each other's
predictions well.

\acknowledgements
I wish to thank the referee for constructive comments that
improved the quality of the paper, and Stacy McGaugh
for providing his data on the mass discrepancy - acceleration
relation.
GG is a
postdoctoral fellow with the National Science Fund (FWO-Vlaanderen).

\label{lastpage}

\begin{thebibliography}{99}

\bibitem[\protect\citeauthoryear{Begeman et al.}{1991}]{B:91}
Begeman, K. G., Broeils, A. H., Sanders, R. H., 1991, MNRAS, 249, 523
\bibitem[\protect\citeauthoryear{Bekenstein}{2004}]{B:04} Bekenstein, J. D., 2004, Phys. Rev. D., 70, 083509 
\bibitem[\protect\citeauthoryear{Bosma}{1981}]{B:81} Bosma, A., 1981, AJ, 86,
  1791
\bibitem[\protect\citeauthoryear{Brada \& Milgrom} {1995}]{BM:95} Brada, R., \& Milgrom, M., 1995, MNRAS, 276, 453
\bibitem[\protect\citeauthoryear{Broeils}{1992}]{B:92} Broeils, A. H., 1992, A\&A, 256, 19
\bibitem[\protect\citeauthoryear{de Blok \& McGaugh}{1998}]{dB:98} 
de Blok, W.~J.~G., McGaugh, S.~S.\ 1998, ApJ, 508, 132 
\bibitem[\protect\citeauthoryear{de Blok, McGaugh \& Rubin}{2001}]{dB:01} de Blok, W. J. G., McGaugh, S. S., Rubin, V. C., 2001, AJ, 122, 2396
\bibitem[\protect\citeauthoryear{de Blok \& Bosma}{2002}]{dBB:02}
 de Blok, W. J. G., Bosma, A., 2002, A\&A, 385, 816
\bibitem[\protect\citeauthoryear{de Jong \& Bell}{2006}]{dJB:06}
de Jong, R. S., Bell, E. F., 2006, preprint, astro-ph/0604391
\bibitem[\protect\citeauthoryear{Donato et al.}{2004}]{Do:95} Donato, F.,
  Gentile, G., Salucci, P.\ 2004, MNRAS, 353, L17 
\bibitem[\protect\citeauthoryear{Famaey \& Binney}{2005}]{F:05} Famaey, B., Binney, J., 2005, MNRAS, 363, 603
\bibitem[\protect\citeauthoryear{Famaey et al.}{2007}]{F:07} Famaey, B., Gentile, G. , Bruneton, J.-P., Zhao, H. S., 2007, Phys. Rev. D, 75, 063002
\bibitem[\protect\citeauthoryear{Gentile et al.}{2004}]{Ge:04} Gentile, G.,
  Salucci, P., Klein, U., Vergani, D., Kalberla, P., 2004, MNRAS, 351, 903
\bibitem[\protect\citeauthoryear{Gentile et al.}{2005}]{Ge:05}
Gentile, G., Burkert, A., Salucci, P., Klein, U., Walter, F., 2005, ApJ, 634, L145
\bibitem[\protect\citeauthoryear{Gentile et al.}{2007}]{Ge:07}
Gentile, G., Salucci, P., Klein, U., Granato, G.~L.\ 2007a, MNRAS, 375, 199 
\bibitem[\protect\citeauthoryear{Gentile et al.}{2007}]{Geb:07}
Gentile, G., Tonini, C., Salucci, P., 2007b, A\&A, 467, 925 
\bibitem[\protect\citeauthoryear{Gentile et al.}{2007}]{Gec:07}
Gentile, G., Famaey, B., Combes, F., Kroupa, P., Zhao, H.~S., Tiret, O.\ 2007c, 
A\&A, 472, L25
\bibitem[\protect\citeauthoryear{Hoekstra, van Albada \& Sancisi}{2001}]{Hoe:01}
Hoekstra, H., van Albada, T. S., Sancisi, R., 2001, MNRAS, 323, 453
\bibitem[\protect\citeauthoryear{Kent}{1987}]{Kent:87}
Kent, S.~M.\ 1987, AJ, 93, 816 
\bibitem[McGaugh \& de Blok(1998)]{1998ApJ...499...41M} McGaugh, S.~S.,  
de Blok, W.~J.~G.\ 1998, ApJ, 499, 41 
\bibitem[\protect\citeauthoryear{McGaugh}{2004}]{M:04} McGaugh, S. S., 2004, ApJ, 609, 652 
\bibitem[\protect\citeauthoryear{McGaugh}{2005}]{M:05} McGaugh, S. S., 2005, ApJ, 632, 859
\bibitem[McGaugh et al.(2007)]{2007ApJ...659..149M} McGaugh, S.~S., de 
Blok, W.~J.~G., Schombert, J.~M., Kuzio de Naray, R., Kim, J.~H.\ 2007, 
ApJ, 659, 149  
\bibitem[\protect\citeauthoryear{Milgrom}{1983}]{Mi:83} Milgrom, M., 1983, ApJ, 270, 365
\bibitem[\protect\citeauthoryear{Milgrom}{1988}]{Mi:88} Milgrom, M.\ 1988, ApJ, 333, 689 
\bibitem[Milgrom \& Sanders(2003)]{2003ApJ...599L..25M} Milgrom, M.,  
Sanders, R.~H.\ 2003, ApJ, 599, L25 
\bibitem[\protect\citeauthoryear{Milgrom \& Sanders}{2005}]{Mi:05} Milgrom, M., Sanders, R. H. , 2005, MNRAS, 357, 45
\bibitem[Moffat(2006)]{2006JCAP...03..004M} Moffat, J.~W.\ 2006, Journal of Cosmology and Astro-Particle Physics, 3, 4 
\bibitem[\protect\citeauthoryear{Moore et al.}{1999}]{Mo:99} Moore, B., Quinn, T., Governato, F., Stadel, J., Lake, G., 1999, MNRAS, 310, 1147
\bibitem[\protect\citeauthoryear{Navarro et al.}{1996}]{NFW:96} Navarro, J. F., Frenk, C. S., White, S. D. M., 1996, ApJ, 462, 563
\bibitem[\protect\citeauthoryear{Navarro et al.}{2004}]{Na:04}Navarro, J. F., Hayashi, E., Power, C., Jenkins, A. R., Frenk, C. S., White, S. D. M., Springel, V., Stadel, J., Quinn, T. R., 2004, MNRAS, 349, 1039
\bibitem[Papadopoulos et al.(2002)]{2002ApJ...579..270P} Papadopoulos, P.~P., Thi, W.-F., \& Viti, S.\ 2002, ApJ, 579, 270 
\bibitem[\protect\citeauthoryear{Persic, Salucci \& Stel}{1996}]{Pe:96} 
Persic, M., Salucci, P., Stel, F., 1996, MNRAS, 281, 27
\bibitem[Pfenniger et al.(1994)]{1994A&A...285...79P} Pfenniger, D., Combes, F., \& Martinet, L.\ 1994, A\&A, 285, 79 
\bibitem[Richtler et al.(2008)]{2008A&A...478L..23R} Richtler, T., Schuberth, Y., Hilker, M., Dirsch, B., Bassino, L., \& Romanowsky, A.~J.\ 2008, \aap, 478, L23 
\bibitem[Salucci \& Gentile(2006)]{2006PhRvD..73l8501S} Salucci, P., Gentile, G.\ 2006, Phys. Rev. D, 73, 128501 
\bibitem[\protect\citeauthoryear{Sancisi}{2003}]{S:03}
Sancisi, R. 2003, in IAU Symp. 220, Dark Matter in Galaxies (Dordrecht: Kluwer), 192
\bibitem[\protect\citeauthoryear{Sanders}{1996}]{S:96}
Sanders, R.~H.\ 1996, ApJ, 473, 117 
\bibitem[\protect\citeauthoryear{Sanders \& McGaugh}{2002}]{S:02} Sanders, R. H., McGaugh, S. S.,
2002, ARA\&A, 40, 263
\bibitem[Sanders \& Noordermeer(2007)]{2007MNRAS.379..702S} Sanders, R.~H., Noordermeer, E.\ 2007, 
MNRAS, 379, 702 
\bibitem[Swaters(1999)]{1999PhDT........27S} Swaters, R.~A.\ 1999, Ph.D.~Thesis,  University
of Groningen
\bibitem[Verheijen \& de Blok(1999)]{1999Ap&SS.269..673V} Verheijen, M.,  
de Blok, E.\ 1999, Ap\&SS, 269, 673 
\end{thebibliography}
\end{document}